# AN EXAMINATION AND REPORT ON POTENTIAL METHODS OF STRATEGIC LOCATION-BASED SERVICE APPLICATIONS ON MOBILE NETWORKS AND DEVICES

Kasra Madadipouya[1]

[1]Department of Computing and Science, Asia Pacific University of Technology & Innovation

*ABSTRACT*

*Mobile technologies are growing significantly in past few years. Many new features and enhancement have implemented in mobile technologies in both software and hardware aspects. Nowadays, cell phones are not just only use for making calls or sending text messages, however, technologies behind mobile phones expanded vastly which facilitate them to offer various types of services. Location-based service is one of the most popular mobile technologies which equipped in new generation of hand phones. The main focus of this paper is to review various strategic location-based applications on mobile networks and devices.*

## 1. INTRODUCTION

Nowadays with the rapid development of information and telecommunication technologies which are integrated in devices, identifying location of mobile devices has become important. Mobile devices are used in various areas like m-banking, m-government and also m-learning [1]. Several technologies with various accuracy and cost such as Geographical Information Systems (GIS), Global Positioning Systems (GPS) are also integrated in mobile devices which are used to determine the location of people, cars, etc. Some most recent location sensing technology based on ultra wideband radio can even achieve accuracies on the order of centimeters in an indoor environment [2].

In addition, the past decade has seen advances in wireless network technologies and an explosive growth in the diversity of portable devices like personal digital assistants (PDAs), tablets, and smart phones [3].Growth of mobile network technologies allows people to access to the Internet where ever and whenever they desire. At the same time, new ways in which consumers use the mobile Internet and in which mobile service applications have emerged, breaking the traditional dominance of mobile network carriers over service delivery [4][5], and leading to a more modular market environment with competing platform models [6]. With mobile devices getting more widespread and technically advanced, the importance of such application is increasing daily.

This paper focuses on reviewing potential methods which could be utilized to develop strategic applications on mobile networks. In the first part of the paper a brief explanation about popularity of mobile phones and importance of them provided. In second part we present complete and comprehensive survey of various applications of location-based services. Then in the last section we conclude the work with its overall outcome.

## 2. SURVEY ON LOCATION-BASED SERVICE APPLICATIONS

The term Location-based services refer to set of applications which can identify the location of a mobile device and offer value added service to the mobile user base on his/her location. Value added services include emergency services, navigation services, entertainment services and such services and such services are increasingly being integrated in various combinations, and to various extents, to meet both user and regulatory requirements in a variety of operational environments. Mentioned applications are act in two ways with the subscriber. The first one is "push" mode where services are pushed to the user automatically without his/her permission. The obvious example of the push mode is advertising application. By contrast in the pull (second) mode the user has this freedom to choose deliverable services from LBS providers based on his needs [7], which tourism application is the common example of pull mode LBS.
LBS are an intersection of three technologies (Figure 1) [8]. It is based on new information and communication technology (NICTS) plus Geographic Information Systems (GIS) with special databases [8].

In order to use a location-based service five basic components are requested. The first element is a mobile device which a tool for the user to receive the needed information and interact with application. The next component is communication which transfers data between a user and a service provider. In other word, the communication network is responsible for interactions among them. The third one is positioning component and it is needed for the processing the user geographical location.

The user location can be acquired by using GPS (Global Positioning System), mobile communication networks and the Internet. Service and application provider is the fourth part of the LBS. It offers a number of different services to the user and responsible for the processing of the requested service [8]. The last component is data and content provider. Service providers mostly do not store all of information which may needed by users. Hence, geographical location of the user and data related the user position is mainly gained from maintaining authority or business and industry partners such as traffic companies, mapping agencies.

- Location-based services provide many advantages to the users and service providers. Some of them are listed below:

- It helps to offer information based on user request among lot of data available on the Internet.

- Service providers with providing relevant information for users assist them to speed up their decisions and activities.

- It decreases the volume of user input data for accessing a service. (LBSs acquire information about users' location automatically via smart mobile devices).

- With sharing location-tagged information, more and up to date data is available for all users.

## 2.1. LBS in emergency services

Emergencies and disasters have been inevitable parts of human life since the beginning of the world. Wide ranges of potential known and formerly unknown hazards keep challenging society even nowadays. Emergency location based-services have been recommended, tested or utilized to overcome impacts of catastrophes. In this scenario LBS is used to determine the exact geographical location [9]. In these conditions LBS can find an individual place after the cellphone user has made an emergency call or a distress short message service request for help who does not know about his/her current location or cannot reveal it due to an emergency circumstance (injury, criminal attack, and etc.)

As a solution LBS also can be utilized to deliver warning notifications and emergency alert information. For instance, emergency information to disaster areas could be broadcast via the new 3G standard "Multimedia Broadcast Multi-cast Service (MBMS)" with rich multimedia content like voice instructions and evacuation maps [10]. Using LBS reduces the side effects and of the emergency and unpredictable disaster. In addition broadcasting alerts in dangers areas can prevent human damages.

For instance, Australia government has commenced to examine its national mobile alert system [10]. Australia authority broadcasts warnings and safety notifications to the public by using methods like mobile phone calls, landlines, emails and short message service (SMS). [11] concluded that SMS and email alerts where considered by the Victorian State Government for the purpose of geographically targeting people in specified areas with information about terror attacks or natural disasters. A warning system for Sydney city was suggested by the New South Wales Premier which residents could join to real time Government SMS and email air pollution health alerts. The system came in response to key recommendations from a Parliamentary air quality inquiry in 2006, which 1600 people were dying annually due to related air pollution disease in New South Wales. The project was done under the Department of Environment control [9].

In addition to the aforementioned example, the New South Wales Government also suggested an electronic warning system which should have allowed ESOs (Police, Fire and Ambulance services) to broadcast SMS messages to all cell phones in emergency target zones in the State [11]. A concept was identify in design phase that the system's need to be operable in all Australian telecommunication networks. Therefore, it could provide evacuation information, safety notifications and alternative roads to stay away the emergency area.

## 2.2. LBS in navigation

Navigation services have been the core of location-based services for many years [12]. In this service, the location of mobile users or vehicles are determined to offer value added services regarding traffic jams, close points of interest (POI) such as gas stations or parking spaces [13]. In addition, navigation system is able to compute the shortest path for the user's destination based on his/her current location [14]. Users can also use the navigation service to find their direction and their current location [8].

Information can be provided for the device via system update or the internet. In the second mode, data is loaded from the internet with one of the various positioning technologies [13]: cellular network (GSM, UMTS) or satellite (GPS, A-GPS).

### 2.3. LBS in information services

Information service is, identifying the user's location and offering information based on his/her location such as list of restaurants near the current place of the user or call a cap from the nearest taxi station, etc. Information services can be used in tourist industry as well. Visitors can utilize the service in their cell phones to easily get information like location of tourist attraction, transportation facilities, accommodation places, medical facilities. A report from [15] showed that around 74 per cent of smart-phone users use their phones for getting real-time location-based information services.

Location-based information also available in another form which is called location-sensitive information services. It is refers to distribution of proper information based on device location, user's behavior and time [8].

### 2.4. LBS in tracking services

Tracking is finding the location of entities such as personal and staff security, goods in transit and vehicles. Initially, it is used to track employee locations. Today, tracking service is used in business area and home life as well [16]. In business area delivery companies could use tracking service to know where their delivery vehicles are. With using tracking service companies can ensure about that all employees are where they should be during working hours. In addition, they can let customers know about the location of a truck that is due to deliver to their house [16].

In home life tracking service allows parents to track their children at anytime and anywhere without any age limitation. Different devices such as GPS receivers, phone applications and car trackers could be used to tracking children with or without their awareness. They are always able to trace the exact location of their kids and monitor activities they are involved in [17]. Some virtual boundaries such as around a school or home can be set up by parents that could alert them when the device arrives or leaves at the zone and moving speed [18]. In addition, the application is applicable in different form for elderly people and patients as well.

### 2.5. LBS in billing services

The service offers calling plans to hand phone users based on their location which they make calls. Lowe cost calls are provided for the users who have unique geographic zones such as home, office or other desired locations. When an LBS user makes or receives a call, LBS checks whether the subscriber is in the preferential calling zones or not. Information is made to the billing system if the subscriber is in one of preferred areas and he/she is charged for the call with preferred rate.

Zones are managed by the carrier's Customer Care with the LBS Web interface. If it is set properly, users can choose their preferred areas from their cellphones or via the internet.
The LBS administrator can dedicate various radiuses for each rate area and exclude some zones in market scopes. In the exclusion zones subscribers are banned to create rate zones. For example, high-income domains or areas with low network coverage [19].

### 2.6. LBS in entertainment services

Entertainment services in LBS are full of entertainment functions. These services can fulfill the recreation needs of the users. In the present study, two services have been classified into this category: the LBS community and LBS games [20].
**2.6.1. Gaming**

LBS gaming contain local factors that common mobile games do not have. image-based services provide games and images related to specific locations. Through solving riddles in contents, users may arrive at specific locations. For instance, Vibo telecom, Inc1. Combined LBS with WAP games for LBS gaming. In Ireland, Vodafone has promoted bot fighter. Orange has also promoted its zone master games in Denmark [20].

### 2.6.1. LBS community/friend-finder

The LBS community has a lot in common with the Internet community. The key difference is that the LBS community contains local images and stories. The user may select the friend from the list, and then search the location of the friend under the category. Thus, mobile location-based services make it a lot easier for friends to meet without the traditional voice calls. In another situation in which friends do not know the place they are supposed to meet, LBS community/Friend-finder service helps solve the problem by verifying locations. Aside from the tracking feature, mobile location-based services also allow users to post events, photos, and diaries on communities. The diaries will show the uploading location and time automatically; therefore, friends in the community can share their diaries. For example, U.S Virgin Mobile and Buzzd in 3rd of September 2008 announced their partnership which Buzzd services will be brought to all Virgin Mobile cellphones with the no-annual WAP contract that provides an interactive and real-time mobile service for Virgin Mobile customers.

### 2.7. LBS in advertising services

Mobile advertising is one of the first applications of LBS because of its potential to earn profit as well as its linkage with mobile commerce activates [21]. Additionally, mobile advertising has achieved significant attention due to its unique features such as personalization which offers new chances to advertisers to put efficient and effective promotions on mobile platform. Several mechanisms can be used to implement mobile advertising along with MLS. Common forms of mobile advertising include broadcasting short messages, mobile banners and proximity-triggered advertisements [21]. Since advertising service has intrusive nature, mostly users have to register for getting such services, perhaps in exchange for other benefits (for instance, special offers or calls rates reduction).

## 3. CONCLUSION

Mobile devices and mobile phones are vastly used and it seems that mobile technologies are vastly utilized in any industry section for various purposes. The paper mainly focused on reviewing different mobile methods with proper case studies in details. Expressed methods can be used in various areas such as business, emergency, entertainment, etc. Various location-based service applications have been covered in this paper which included emergency services, navigation and traffic flow, information services, tracking services, billing services, entertainment services, and advertising services. In emergency area location-based service application can be utilized in disaster management to inform people about hazard areas by sending notifications. In navigation and traffic flow with use of LBS application, a way of traffic management and controlling roads and routes traffic can be established in order to assist drivers. LBS application also can be useful in business point of view via increasing customer satisfaction with providing various services such as billing, advertising and entertainment services.

## REFERENCES

[1] Ghadirli, H., & Rastgarpour, M. (2012, September). An Adaptive and Intelligent Tutor by Expert Systems for Mobile Devices. In International Journal of Managing Public Sector Information and Communication Technologies, 3(1).

[2] Ververidis, C., & Polyzos, G. (2002, July). Mobile marketing using a location based service. In Proceedings of the First International Conference on Mobile Business. Prentice-Hall.

[3] Hurson, A. R., & Gao, X. (2009). Location-Based Services. Encyclopedia of information science and technology, 2456-2461.

[4] Gerum, E., Sjurts, I., & Stieglitz, N. (2004). Industry convergence and the transformation of the mobile communications system of innovation. In ITS 15th Biennial Conference, Berlin, Germany.

[5] Reserve, F. (2012). Board of Governors of the Federal Reserve System. Consumers and mobile financial services.

[6] Ballon, P., Walravens, N., Spedalieri, A., & Venezia, C. (2008). The reconfiguration of mobile service provision: towards platform business models. Available at SSRN 1331549.

[7] Ververidis, C., Polyzos, G. C., & Mehdi, K. P. (2006). Location-Based Services in the Mobile Communications Industry. Encyclopedia of E-Commerce, E-Government and Mobile Commerce, Idea Group Reference, Hershey, USA.

[8] Steiniger, S., Neun, M., & Edwardes, A. (2006). Foundations of location based services. Lecture Notes on LBS, 1, 272.

[9] Aloudat, A., & Michael, K. (2011). The application of location based services in national emergency warning systems: SMS, cell broadcast services and beyond.

[10] Aloudat, A., Michael, K., & Yan, J. (2007). Location-based services in emergency management-from government to citizens: Global case studies. Faculty of Informatics-Papers, 562.

[11] Michael, K., Abbas, R., Aloudat, A., & Al-Debei, M. (2011). The value of government mandated location-based services in emergencies in Australia. Journal of Information Technology Research, 4(4), 41-68.

[12] Rao, B., & Minakakis, L. (2003). Evolution of mobile location-based services. Communications of the ACM, 46(12), 61-65.

[13] Greßmann, B., Klimek, H., & Turau, V. (2010, March). Towards ubiquitous indoor location based services and indoor navigation. In Positioning Navigation and Communication (WPNC), 2010 7th Workshop on (pp. 107-112). IEEE.

[14] Hand, A., Cardiff, J., Magee, P., & Doody, J. (2006). An architecture and development methodology for location-based services. Electronic Commerce Research and Applications, 5(3), 201-208.

[15] Zickuhr, K. (2012). Three-quarters of smartphone owners use location-based services. Pew Internet & American Life Project.

[16]     Brown, A. K., & Sturza, M. A. (1995). U.S. Patent No. 5,379,224. Washington, DC: U.S. Patent and Trademark Office.

[17]     Applewhite, A. (2002). What knows where you are?. Pervasive Computing, IEEE, 1(4), 4-8.

[18]     Burke, L. (2012). Parents tracking children with GPS. Yahoo News Australia. Retrieved from http://au.news.yahoo.com/thewest/a/-/newshome/13495616/parents-tracking-children-with-gps

[19]     Chan, N., & Lars, H. (2003). Introduction to Location-Based Services. Lund University GIS Centre.

[20]     Chen, P.T. & Lin,Y.S. (2011). Mobile Location-Based Services: An Empirical Study of User Preferences.

[21]     Giaglis, G. M., Kourouthanassis, P., & Tsamakos, A. (2003). Towards a classification framework for mobile location services. Mobile commerce: technology, theory, and applications, 67-85.

Kasra Madadipouya was born on 25th of July 1989. He received the B.S degree in Computer Software Engineering from Shahid Shamsipour University in Tehran, Iran 2011. M.S degree in Software Engineering from  Asia Pacific University of Technology & Innovation in Kuala Lumpur, Malaysia 2013. Currently he is working with professional software development in Malaysia

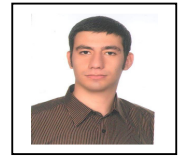